\documentstyle[epsf]{l-aa}

\newcommand{\km}{\,\mbox{km}\,\mbox{s}^{-1}}

\begin{document}

\thesaurus{03(11.09.1; 11.09.4; 11.11.1)}

\title{ Strong non-circular motions of gas in the spiral  galaxy NGC~1084
\thanks{Based on observations collected with the 6m and 1m telescopes
of the Special Astrophysical Observatory (SAO) of the
Russian Academy of Sciences (RAS), operated under the
financial support of the Science Department of Russia (registration number
01-43).}
}
\author{A.V. Moiseev}
\institute{Special Astrophysical Observatory , Nizhnij Arkhyz,
Karachaevo-Cherkesia, 357147, Russia (moisav@sao.ru)}
\date{Received ...../ Accepted .....}
\maketitle

\begin{abstract}
The results of $H_\alpha$ and [NII]$\lambda 6583$
observations of the spiral galaxy NGC 1084 at the SAO 6m telescope with
a Fabry-Perot interferometer are presented. The complex structure of
the emission line profile has revealed  the presence  of giant
star formation regions (``spur'') in the NE part of the galaxy. In
this region the $H_\alpha$ line shows two distinct components, with
line-of-sight velocity differences of $\pm(100-150)\km$.
 The first component corresponds to  normal circular rotation.
The  second velocity component may be a signature  of non-circular
motions of the ionized gas in shock wave fronts. An increase of the
$\mbox{[NII]}/H_\alpha$ line ratio confirms the shock wave
interpretation of these features.
The ionized gas kinematics in this galaxy is discussed.

\keywords{galaxies: individual: NGC 1084 -- galaxies: ISM --
          galaxies: kinematics and dynamics}

\end{abstract}

\section{Introduction}

A velocity field of interstellar gas in spiral galaxies  is not only a
 good indicator of mass distribution and structural properties of
 their disks, but can also reveal perturbations in diffuse matter
 related to local sources of energy. Non-circular motions of gas can
locally trigger an active star formation, and, in turn, may be the result of
 the collective action of massive young stars on a short
 time scale.

Actually, a model of circular gas motion
in the disk of any galaxy is no more than a first approximation to
the real kinematic picture. Apart from the evident cases of tidal
forces or of an active
nuclear region (which will not be considered here), responsible for
peculiar gas motions, the most common reason of deviation from circular
motion is a presence of a spiral density wave or a bar. In these
cases, non-circular velocity components have   well-ordered systematic
character related with the optically observed structure.

However in some galaxies local velocity
perturbations exceeding $30-50\km$ have been discovered, which cover large
regions from a few hundred pc up to a few kpc size. Good
examples are  two regions observed in M 101 in the HI line (van der
Hulst \& Sancisi \cite{hulst}) which show
peculiar velocity components reaching $150\km$, as well as giant HI
supershells expanding with velocities of $25-45\km$ in NGC 4631
( Rand \& van der Hulst \cite{rand}),  NGC 1313
(Ryder et al. \cite{ryder}) and IC 2574 (Walter et al. \cite{walter}).
The typical kinetic energy of
perturbed gas motions in these cases is about $10^{53} - 10^{54}$
ergs, and their connection with  sites of star formation is
obvious. Although in different galaxies the nature of local velocity
peculiarities may  not be the same, there are two  ways to
explain it: local bursts of star formation (stellar winds, explosions
of supernova or hypernova) or accretion of intergalactic gas
clouds and dwarf gas-rich galaxies (see references and discussion in
Rand \& van der Hulst \cite{rand}).

In this paper we describe the discovery of an extended region of
unusual strong non-circular motion of gas in the spiral galaxy NGC
1084 from optical observations in the $H_\alpha$ and [NII] emission lines.

NGC 1084 is a late-type spiral galaxy  classified as
SA(s)c in the Reference catalog of bright galaxies (RC3). The
distance  adopted in this paper is 18.5 Mpc
($H_0=75\,\mbox{km}\,\mbox{s}^{-1}\mbox{Mpc}^{-1}$). At  first glance, NGC 1084 is
a normal galaxy with mildly inclined disk, a regular two-armed
grand design spiral structure, and without close optical companions
or any morphological peculiarities. The rotation of the gas in this
galaxy was measured on several occasions.
Burbidge et al. (\cite{burb})  obtained three long-slit spectra
in the spectral range near $H_\alpha$,
making two cuts along the major axis and one cut along the minor axis.
Yet, the  accuracy of their measurements
was low, and the obtained velocity curve was  unreliable.
Kyazumov (\cite{kyaz}) has performed a long-slit study of NGC 1084.
He has obtained improved  line-of-sight velocity distributions,
although the shape of the rotation curve remains
uncertain. The  maximum  rotation velocity of  $130-140\km$ which he finds,
has been confirmed  later by Afanasiev et al.
(\cite{afanas}).  These authors have obtained long-slit
spectroscopy with a digital detector (IPCS $512\times 512$)
at  the 6m telescope. The rotation curve of the ionized gas
has been of a higher accuracy than  previous determination. In
particular, it is found that the velocity curve reaches its maximum
very close to the center -- at a radius of $R\approx10''-15''$. Two long-slit
cross-sections --  along the major axis and under an angle of $30^\circ$ to it
-- indicate flat velocity distributions up to  $\pm 50''$ from the
center. Besides, the higher spatial resolution enables them to detect some
non-circular phenomena.
Firstly, in the central region ($R<5''$)  a difference in velocities
measured from $H_\alpha$ and from the forbidden emission lines ([NII] and [SII])
is found. The authors have interpreted it as a possible signature
 of two differently rotating gaseous systems, where non-circular
velocities associated to the forbidden lines would be caused by
a low-contrast nuclear minibar.
Secondly, at $PA=4^\circ$ an extended region located $40''$ to the
N has been localized,  which shows a negative excess of
line-of-sight velocity up to $30\km$. No explanation is  proposed
for this feature.

     In  this  paper,   new  observations  of NGC 1084 with a
Fabry-Perot interferometer at the SAO 6m telescope are presented.
The main goal of the observations was the study of the velocity
field of the ionized gas in the galaxy as a whole.
We  focus on  investigating
strong non-circular gas motions in the northern part of the galaxy.
The analysis of  gas grand-design motions in the spiral structure
will be given in a forthcoming papers.

The paper is structured as follow: In the next section (Sect. 2), we
describe the  observations  and data reduction; the ionized gas kinematics
is described in Sect. 3; possible explanations of nom-circular gaseous
motions are discussed in Sect. 4; conclusions are drawn in Sect. 5.

\section{Observation and data reduction}
\subsection{Observations with the Fabry-Perot Interferometer}

\begin{figure}
\centerline{\epsfbox{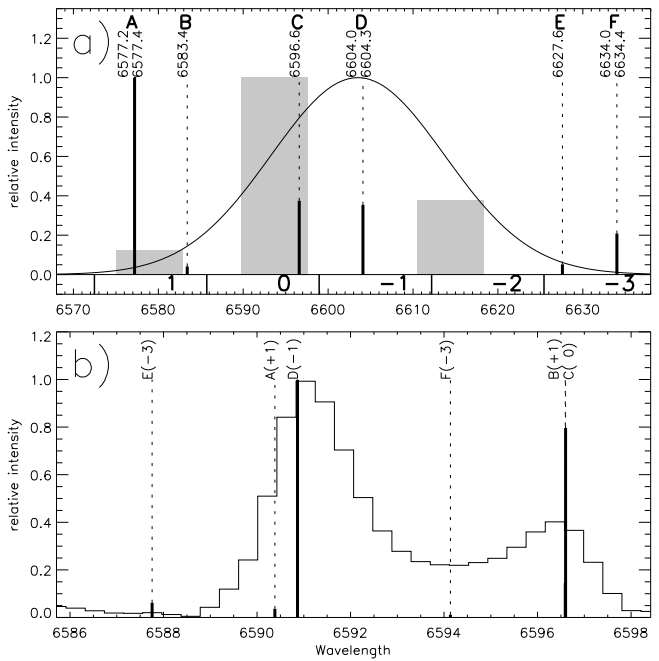}}

 \caption{ The emission lines into the order separating filters.
 {\bf a} -- on the common wavelength scale, the solid gaussian is the filter
transmission, the gray boxes are the emission lines of NGC 1084:
[NII]$\lambda 6548$,  $H_\alpha$ and [NII]$\lambda 6583$ (see text); as well the
lines of the night sky are plotted with their wavelengths  and relative
intensities values from
Osterbrock et al.(1996). Near the wavelength axis the numbers of
the interference orders   relative of the $H_\alpha$
order are indicated.
   {\bf b} -- the night sky spectrum on the $H_\alpha$-order wavelength scale.
 The thin line is the mean of the sky spectrum from our FPI data. The thick lines
 are the night sky lines from the neighboring
 orders. The lines marked with letters are from figure (a),  the
 interference order for each line is given within  brackets.
}
\label{figfilter}
\end{figure}

\begin{figure*}
\centerline{\epsfbox{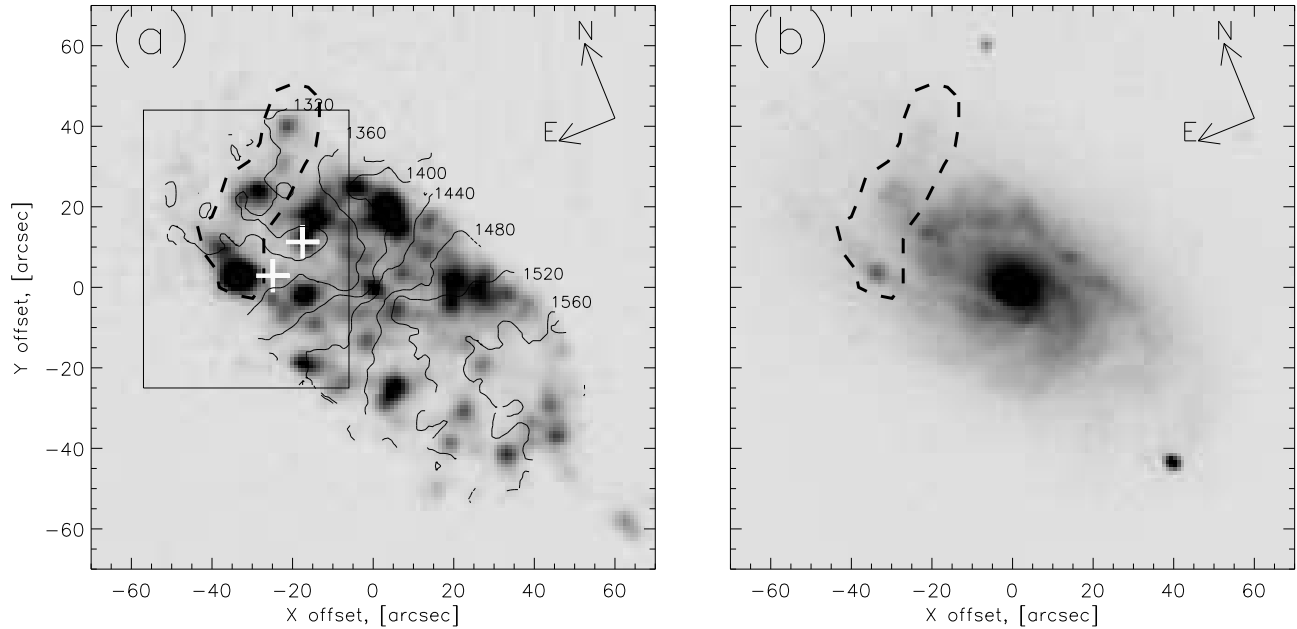}}
 \caption{  Direct images  of  NGC 1084. The
 dash  contour  shows  the  ``spur''  region (see text).  {\bf a} - $H_\alpha$
 image from FPI-data.  White  crosses mark the locations of SN 1996an and SN
  1998dl.  The  rectangular  region corresponds to the
  field displayed in  Fig. 3 and Fig. 4. Contours of the
  $H_\alpha$ velocity field are superposed. Labels indicate the
  velocity values in   $\km$.
   {\bf b} - I-band image  obtained  at  the 1m telescope.
 }
\label{figim}
\end{figure*}

The two-dimensional velocity field of NGC 1084 was obtained on
October 26, 1995 at the SAO 6m reflector, using a scanning Fabry-Perot
interferometer (FPI)  installed in  the pupil plane of a
focal reducer  attached to the f/4 prime focus of the telescope.
The detector was an intensified photon counting system (IPCS). The
 observational parameters are given in table \ref{fpi}.

\begin{table}
\caption[]{FPI observations parameters.}
\label{fpi}
\begin{tabular}{ll}
\hline
Number of pixels                  &     $256\times256$ (binning $2\times2$)       \\
Pixel size                        &     0\farcs92                     \\
Galaxy $H_\alpha$ wavelength      &     $\lambda_g=6594$  \AA \\
Etalon interference order         &     501 at $6563$\AA \\
Free spectral range               &     $13.41$\AA~~$(602\km~\mbox{at}~\lambda_g)$ \\
Number of spectral channels       &    32                            \\
Channel size                      &     $0.41$\AA~~$(18.8\km~\mbox{at}~\lambda_g)$       \\
Exposure time for each channel    &    225 s                          \\
Total exposure time               &    7200 s                        \\
Seeing                            &    $2''$                         \\
\hline
\end{tabular}
\end{table}

An order separating filter with $FWHM\approx26$~\AA~~was used, centered at
$6603$\AA, close to the redshifted galactic emission line
$H_\alpha$. The filter bandpass includes also the nitrogen emission line
[NII]$\lambda 6583$. This line  falls into the interfringe of
the etalon - very close to the next order
$H_\alpha$ line. Usually such situation complicates  data processing and
interpretation, but in our case the proximity of
$H_\alpha$ and [NII] interference rings was purposely used to compare  gas
kinematics in two emission lines from the same observational data set.

Observational data were converted into a cube of 32 images.
A neon lamp spectrum was used for phase calibration. Reduction of  the
observational data was performed using the software
ADHOC developed at the Marseille Observatory (Boulesteix \cite{boules}).
It  includes a phase map construction (wavelength
calibration), subtraction of the night sky emission,  spatial and
spectral smoothing. The spatial resolution of  our data, after
smoothing, is 3\farcs5, and the spectral resolution is close to $50\km$.
Uncertainty of  velocity measurements depends mostly on calibration
errors and is about $10\km$.

After phase calibration, the  first spectral channel  corresponds to
6585.5  \AA~~  ($1039  \km$  at  the  redshifted  $H_\alpha$  line).
The [NII]$\lambda6583$ line is observed in the
-2  interference  order  relatively  to the $H_\alpha$-line order
and  has  a visible shift  of $-6.2\AA$ ($285\km$) from the $H_\alpha$ line
position.   Therefore, the  emission lines are certainly
separated since  the spectral resolution
of  the  FPI  is  about $50\km$.

Fig. \ref{figfilter}b shows the transmission curve of the narrowband
order separating filter, while the $H_\alpha$
and  [NII]$\lambda6548,6583$ emission lines positions are marked as gray boxes.
The width of  these boxes  corresponds to  the full range of observable
velocities ($1230-1590\km$). The high velocity components of the $H_\alpha$ line were
included. The relative heights of the gray boxes have been set from normal
emission lines ratio in HII regions.
The flux from the [NII]$\lambda6548$ line must then be 10 times lower
than the flux from the [NII]$\lambda6583$ line  due to  the filter
transmission. Indeed, there is   no traces of [NII]$\lambda6548$ in
our FPI spectra.

Relative intensities of the night sky emission lines from
Osterbrock et al.(\cite{ost}) are shown  in Fig.\ref{figfilter}.
The FPI's mean night sky spectrum was obtained as an integrated emission on the
detector's part  which is free from emission of the galaxy and its ghost
image.  Then the mean night  sky spectrum was subtracted from all
Fabry-Perot spectra.  The total  mean  night  sky spectrum  plotted in
 Fig. \ref{figfilter}b and the individual sky lines from Fig. \ref{figfilter}a
 are superimposed on the FPI spectrum.
  The relative intensities of these emission lines were multiplied on the filter
bandpass transmission curve
and the wavelengths of the night sky lines were converted to the
wavelength scale for  the $H_\alpha$ interference order.

The night sky lines $\lambda6596$ (label C in Fig. \ref{figfilter}) and
$\lambda 6604$ (label D) are the  main contributors to the   observed FPI spectra.
Contribution from other lines is negligible  ( including
$\lambda 6577$ which is the brightest in the filter bandpass but is  located
in the extreme blue wing of the filter). Discrepancies between the FPI night sky
spectrum and the line intensities and positions from Osterbrock et
al.(\cite{ost}) are due to calibration errors and  night sky brightness
variations.

In Sect. 3 it will be shown than all non-circular components of the
object's emission lines are  brighter than  the
mean night sky spectrum. Therefore the errors due to subtraction of the
night sky lines have no influence on the  measurement of the high
velocity motions of the gas.

The velocity map and monochromatic $H_\alpha$
 and red continuum  images of the galaxy were constructed after sky
subtraction and smoothing procedures.
All spectral channels within $\pm150\km$ (8 channels) from the
channel with maximal intensity were summed in each
pixel to obtain the $H_\alpha$ image. The non-circular component of
$H_\alpha$ has a relative velocity larger than $100\km$
only in regions where its intensity is negligible in
comparison with the main $H_\alpha$ component (see below). Therefore
the velocity range $\pm150\km$ is optimal for measuring  the
total flux.   The  $H_\alpha$ flux
is calibrated using the integrated flux of H$_\alpha$+[NII] in NGC 1084
from Kennicutt \& Kent (\cite{kenken}) and assuming a line ratio
$L_{H_\alpha+[NII]}/L_{H_\alpha}$ = 1.5.

Baricenters of the $H_\alpha$ and [NII] lines were calculated to obtain the
full-format velocity fields (the map of the first moment) in these
lines.  Fig. \ref{figim}a. shows  the $H_\alpha$ image of the galaxy
and isovelocities of the $H_\alpha$ velocity field.
For the  regions with complex  emission line profiles we used a
multi-component gaussian analysis.

\subsection{Photometric observations}

Images of the galaxy  were obtained on  January 14, 1997, at the SAO
1m Zeiss reflector with a CCD camera through the V filter of Johnson's
system and R$_C$, I$_C$ filters of Cousin's system.
The pixel size of CCD was 0.49\arcsec,
the seeing was about 2\farcs5. Standard stars from Landolt (\cite{land})
were observed for flux calibration.   The rms error in the
determination of the photometric zero points was 0\fm01.

The  image  of NGC 1084 in  I$_C$ is shown in Fig. \ref{figim}b.
Let us note that the spiral structure which is  well seen in the broad band image
can hardly be traced in $H_\alpha$ (Fig. \ref{figim}a).

\begin{figure}[t]
\centerline{\epsfbox{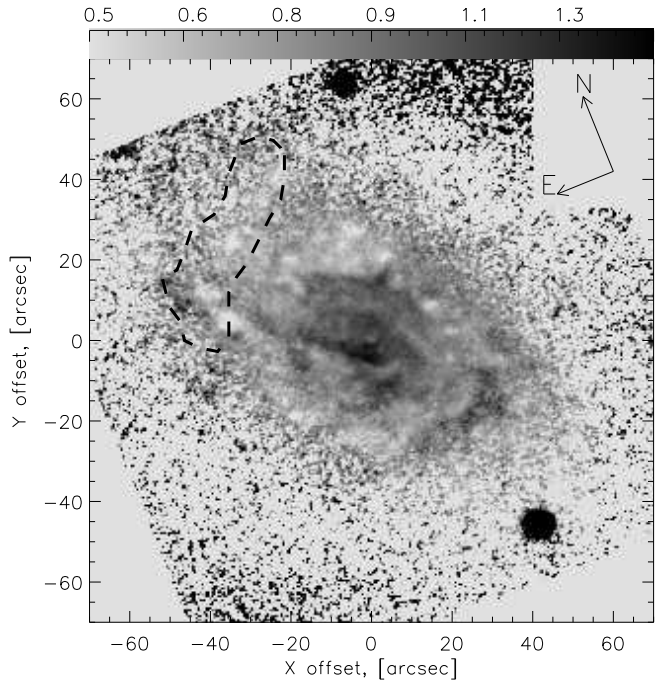}}
 \caption{  The color index (V-I$_C$)  of NGC 1084.
  The  dash  contour  shows  the  ``spur''  region (see text).
 }
\label{figvi}
\end{figure}

A map of the (V-I$_C$) color index is shown in Fig. \ref{figvi}.
A thick clumpy dust lane ("red" region in this image) appears in  the SE
inner part of NGC 1084 suggesting that this side of the galaxy
is the closest to us. Since the SW part of the galaxy is redshifted and the NE
part is blueshifted (see the velocity field on  Fig. \ref{figim}),  the spiral arms are trailing. This situation
is ordinary for spiral galaxies.

\section{Gas kinematic properties}

 In the NE part of the galaxy
there appears a chain of HII regions which does not match the shape
of the spiral arm. This structure, which we call a ``spur'', is
outlined in Fig. \ref{figim} by a dash contour. It begins near a bright
HII region and  extends nearly perpendicular to the
spiral arm. The total $H_\alpha$ luminosity of the  spur reaches
$15-20\, \%$ of the total $H_\alpha$ luminosity of the galaxy.

\begin{figure}[t]
\centerline{\epsfbox{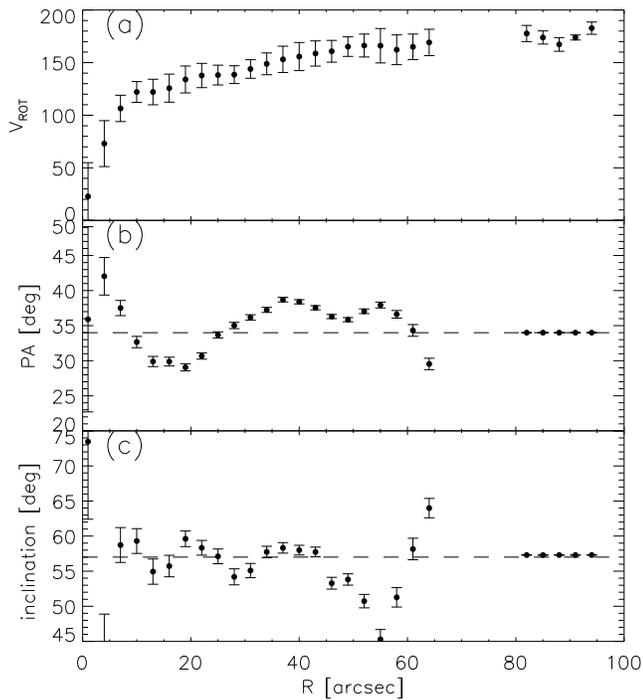}}
 \caption{
Analysis  of the velocity  field of the ionized gas in circular
rotation  approximation:  {\bf a} -- rotation  curve, {\bf b} -- position angle of the
kinematical major  axis, {\bf c} -- disk  inclination. Dashed lines
indicate the mean values of the orientation parameters
 }
\label{figrc}
\end{figure}

The mean velocity curve used as the reference curve of circular
rotation was derived from the velocity measurements across the
entire body of the galaxy by applying the custom developed software
based on  the algorithm
described by Begeman (\cite{begeman}) for pure circular rotation.
As a first step we find the dynamical center position and the mean
value of the
systemic velocity $V_{SYS}$. As a second step these parameters are
fixed, and the   position angle of the kinematical major axis $PA$ and
inclination $i$  are estimated in  tilted rings of 3\arcsec~
width. Fig. \ref{figrc} shows the radial dependence for the
rotation velocity $V_{ROT}$  (Fig.  \ref{figrc}a), for $PA$ (Fig. \ref{figrc}b) and for
disk inclination $i$ (Fig. \ref{figrc}c).
At $R>65''$, velocity data are available only for small emission
islands in the WE (part see Fig. \ref{figim}a) and in this region
we fix the mean values for $i$ and $PA$.
The resulting mean disk parameters
( $i=57^o$,  $PA=34^o$,  $V_{sys}=1435\km$) are in  good agreement with those
found by Afanasiev et al. (\cite{afanas}).
The disk orientation parameters being fixed at their mean value and
the rotation  curve  being  extracted  from the $H_\alpha$ velocity field
(Fig. \ref{figrc}a), this figure shows that the circular rotation velocity is approximately
constant and does not exhibit any peculiarities for the radius range $R=40''-90''$.

In the most part of the galaxy the profiles of the $H_\alpha$ and [NII]
emission lines are quite symmetrical and have a gaussian shape
(except the central region $R<15''-20''$ where a bar-like structure may be
located). But in the ``spur'' the  emission profiles
differ from the common picture.
 In many locations in the ``spur'' the  H$_\alpha$  profiles split into
two components: a ``normal'' component, close to the expected
one from the circular rotation velocity field,  and an ``abnormal'' component,
shifted by  $\pm(50-150)\km$.  To study this peculiarity in detail,
we have binned resulting in   our data cube (by $2\times2$) an enlarged
pixel size of 1\farcs84.
Double-horned profiles of the spectral lines were fitted by two
gaussians, corresponding to the ``normal'' and ``abnormal'' velocity
components.
Both $H_\alpha$ and [NII]$\lambda 6583$ emission lines were used. In some regions of
the spur the latter appears to be strongly enhanced, almost up to the level of
the $H_\alpha$ line intensity.

\begin{figure*}
\centerline{\epsfbox{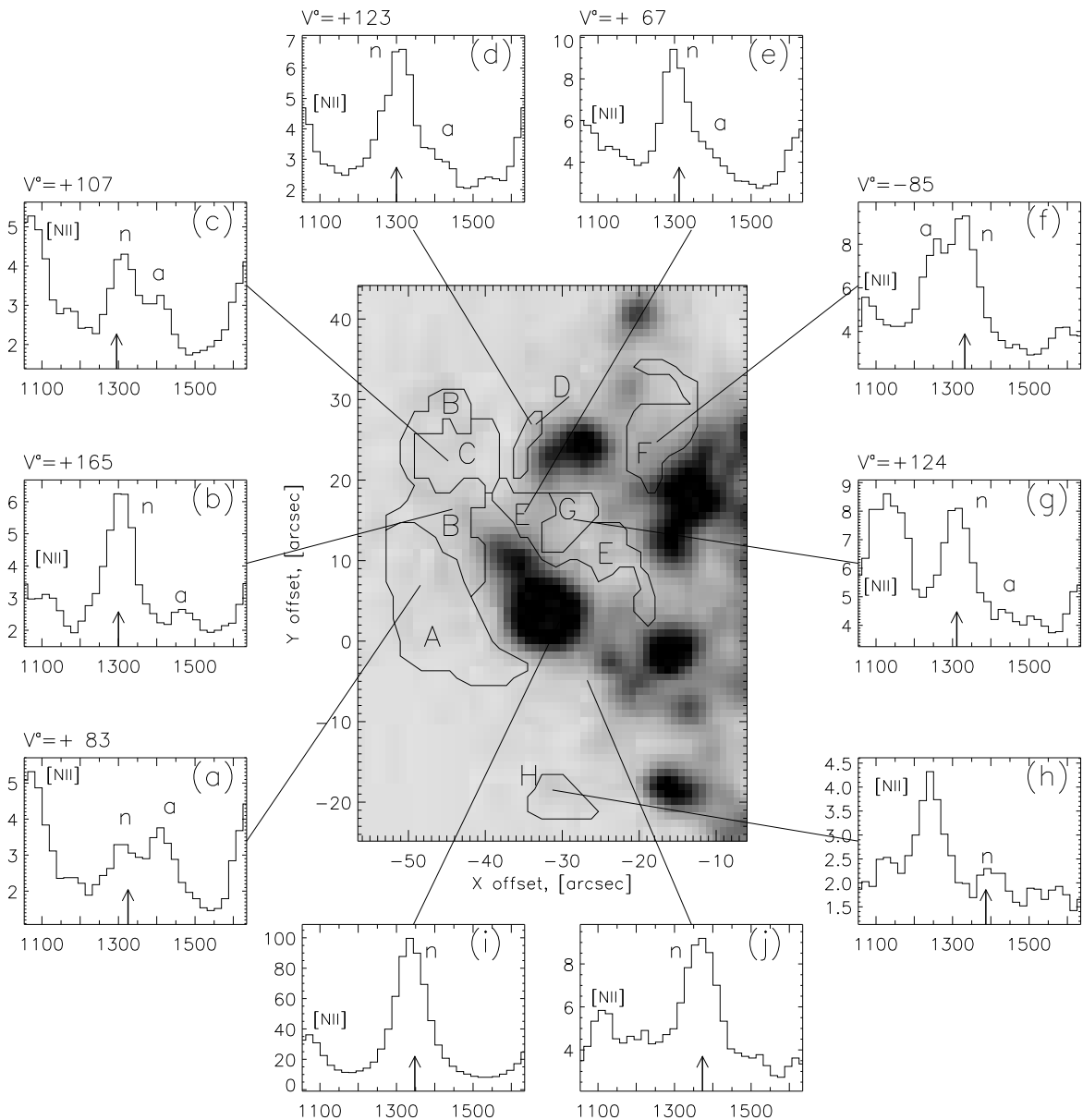}}

 \caption{
$H_\alpha$  image  of  the  ``spur''  and examples of emission
line  profiles from regions around/in the ``spur'' (see the text for
details). In each spectrum  the  x-axis is in $\km$ and the y-axis
is a intensity in relative units . The arrow in each frame
corresponds to the  value  of the  mean  circular rotation
velocity. ``Normal'' and ``abnormal''  components  of  $H_\alpha$
are marked  as  ``n''  and ``a'', [NII]$\lambda 6583$ line is marked
as ``[NII]''. Above each frame the velocity  of  the  ``abnormal''
component of $H_\alpha$ is given (if  present).
}
 \label{figprof}.
\end{figure*}

Fig. \ref{figprof} reproduces the enlarged $H_\alpha$-image of the ``spur'' ,
where different regions are identified by letters A -- H. Typical
line profiles of $H_\alpha$ and [NII]$\lambda 6583$ are also
shown   for   every   region  in  Fig.  \ref{figprof}i  and  Fig.
\ref{figprof}j. The vertical arrows in
each frame corresponds to the velocity of  circular rotation.

The regions with abnormal velocity components have a complex
shape and are located  mainly  between the bright
HII regions of the ``spur'', avoiding the sites of active star formation.
Although the brightness of emission lines away from HII regions is rather
low, the observed anomalies are reliable features.
As an illustration, Fig. \ref{figprof}j shows  normal line
profiles obtained for a low brightness region. The bright giant HII
regions possess quite normal line profiles (see Fig. \ref{figprof}i),
and their velocities correspond to the expected ones for the pure rotation.

\begin{figure}[t]
\centerline{\epsfbox{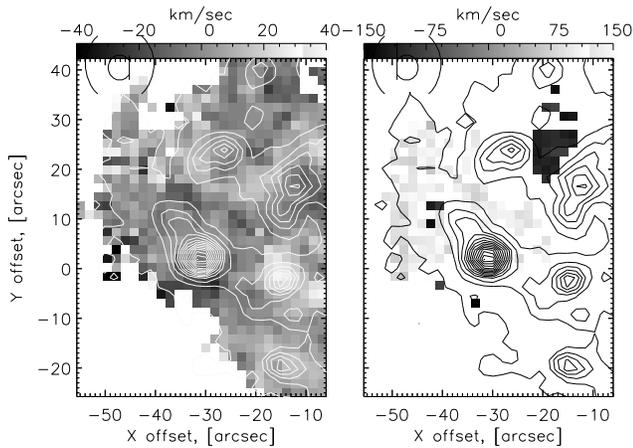}}
 \caption{
 Residual  velocities  in the  ``spur''  after subtraction of a
pure circular rotation,  {\bf a}  --  for the   ``normal''  $H_\alpha$
component and {\bf b} -- for the ``abnormal'' component. $H_\alpha$ intensity contour
is overlapping these maps.
 }
\label{figres}
\end{figure}

To obtain a map of residual
velocities in the given region of the galaxy, the simulated 2D
line-of-sight velocity field corresponding to the mean rotation
curve (Fig. \ref{figrc}a), was subtracted from the observed
velocity field.
The map of the residual velocities overlapped by the isophotes of
the $H_\alpha$ image is shown in Fig. \ref{figres} -- separately for
``normal'' and ``abnormal''  components. It shows that local velocities of
the ``normal'' components are not perturbed by HII regions. They are
in good agreement with the
rotation, and hence  are related to the non-disturbed gas. Let us note
however that the dispersion of residual velocities is about
$20\km$ that exceeds the observational errors and might reflect
velocity perturbation by a density wave. On the contrary,
velocities of the ``abnormal'' components differ from circular velocities
by about $100\km$ and as  mentioned above they are observed
mostly between the bright HII regions.

     Kinematic  and  photometric parameters of the regions marked
in Fig. \ref{figprof}, are given in Table \ref{spur}.
Column (1)   gives the  region identification in
agreement with Fig. \ref{figprof}. Columns (2) -- (3) give the mean
velocity residuals (observed velocity minus circular velocity) for
``normal'' ($V^n$) and ``abnormal'' ($V^a$) velocity components, measured
from $H_\alpha$ profiles. Column (4) gives the residual velocities found
for [NII]$\lambda 6583$, columns (5) and (6) provide the intensity
ratios of [NII] to $H_\alpha$  lines and the ratio of
``normal'' to ``abnormal'' $H_\alpha$ components. The errors
in columns (2)-(6) were
obtained by the intensity-weighted averaging of values over the whole
region. Column  (7) gives the total $H_\alpha$ luminosity (in
$10^{39}\,\mbox{erg~s}^{-1}$).
Line intensities were not corrected for internal absorption. Such a
correction would  increase $L_{H_\alpha}$, but would not change the
intensity ratios.

\begin{table}
\caption{Residual velocities  and
line ratios for
different regions of the ``spur''.}
\label{spur}
\begin{tabular}{crrrrrr}
\hline
region& $V^n$&$V^a$&$V_{[NII]}$ &[NII]/$H_\alpha $&$H_\alpha ^a/H_\alpha ^n$ &$L_{H_\alpha}$\\
(1)   &   (2)  &  (3)  &       (4)    &         (5)    &   (6)     & (7)  \\
\hline
A     &  $-6$  & $  97$& $ 12$        &  $ 1.56$       &  $ 1.17$  & 1.28 \\
      &$\pm3$  & $\pm4$& $ \pm2$      &  $\pm0.09$     &  $\pm0.33$&      \\
B     &$  3 $  & $ 108$& $ 16$        &  $ 0.63$       &  $ 0.28$  & 1.21\\
      &$\pm2$  & $\pm6$& $ \pm2$      &  $\pm0.03$     &  $\pm0.04$&     \\
C     &$  7$   & $ 100$& $ 31$        &  $ 1.12$       &  $ 0.69$  & 0.67\\
      &$\pm3$  & $\pm7$& $ \pm3$      &  $\pm0.05$     &  $\pm0.30$&     \\
D     &$  7$   & $ 144$& $ 18$        &  $ 0.67$       &  $ 0.12$  & 0.25\\
      &$\pm3$  & $\pm8$& $\pm1$       &  $\pm0.07$     &  $\pm0.03$&     \\
E     &$ -7$   & $ 101$& $ 17$        &  $ 0.61$       &  $ 0.19$  & 1.93\\
      &$\pm1$  & $\pm4$& $\pm5$       &  $\pm0.03$     &  $\pm0.02$&     \\
F     &$ -1$   & $-108$& $ 14$        &  $ 0.50$       &  $ 0.35$  & 1.78\\
      &$\pm2$  & $\pm3$& $\pm6$       &  $\pm0.02$     &  $\pm0.03$&     \\
G     &$ -1$   & $ 122$& $ 54$        &  $ 0.87$       &  $ 0.18$  & 0.63\\
      &$\pm3$  & $\pm5$& $\pm8$       &  $\pm0.06$     &  $\pm0.03$&     \\
H     & ---   &    ---& $-15$        &  $ 4.7$        &   ---    & 0.27\\
      &        &       & $\pm4$       &  $\pm2$        &           &     \\
\hline
\end{tabular}
\end{table}

As seen in Table \ref{spur}, the ``abnormal'' component is especially
strong on the periphery of the ``spur'' (regions A and D).
It is just where the  relative intensity of the
nitrogen line is observed to be the largest: [NII]$\lambda 6583$ in these regions is
comparable
to H$_\alpha$ and sometimes is larger (see Fig. \ref{figprof}a and
\ref{figprof}c).

It should be noted however that there is an uncertainty in the  estimates of line
ratios due to continuum subtraction
 the  overlapping of two  interference orders. In addition, the
transparency of the interference filter is different for
[NII]$\lambda 6583$ and H$_\alpha$ lines, and the velocity variations
 of these components may also change their observed
relative intensity. But it cannot affect the results
significantly because within the same region the observed
velocity  range  of any component usually does not exceed $50\km$.
Note also that
independent measurements of the line intensity ratios in bright HII regions
of the ``spur'' carried out at the same telescope with the long
slit  spectrograph  UAGS  (A.N.  Burenkov, private communication)
give  [NII]$\lambda  6583$/H$_\alpha\approx0.31\pm0.03$,  which is
in good agreement with  our own measurements.

In the Region B and in the Region D which captures  the
extension of the bright  HII region, the [NII] line intensity is relatively low
([NII]$\lambda 6583$/H$_\alpha \approx 0.5 $), and the non-circular
component is seen only as an asymmetry in the $H_\alpha$
profile. In  Region E, which extends over about 2 kpc
between two bright HII regions, the relative intensity of the
``abnormal'' component is also low, less than 20\% of the normal
one. Non-circular
motions of the  gas are traced by an enhanced ``red'' wing of the
$H_\alpha$ profile. A similar asymmetry is typical for [NII] line profiles in
this region. In the Region G, neighboring E, the intensity of
[NII] becomes  comparable to $H_\alpha$. Non-circular components of
$H_\alpha$ are not detected (Fig. \ref{figprof}g), but the [NII] line is
redshifted by at least  $50\km$ with respect to $H_\alpha$.
The situation is different in the isolated Region F, lying at the
inner edge of  the ``spur''. The  relative intensity of [NII] looks normal
here, but the $H_\alpha$ line possesses a bright blue-shifted non-circular
component. Note that this is the same region where the
negative relative velocity excess was found earlier by Afanasiev et
al.(\cite{afanas})
from long-slit observations of $H_\alpha$ with lower
spectral resolution.

Finally, the Region H, lying on the continuation of the ``spur''
differs from the other regions by an unusually weak $H_\alpha$ line
([NII]$\lambda 6583$/H$_\alpha \approx 5 \pm 2 $) and by the absence
of a noticeable non-circular component.

\begin{figure}[t]
\centerline{\epsfbox{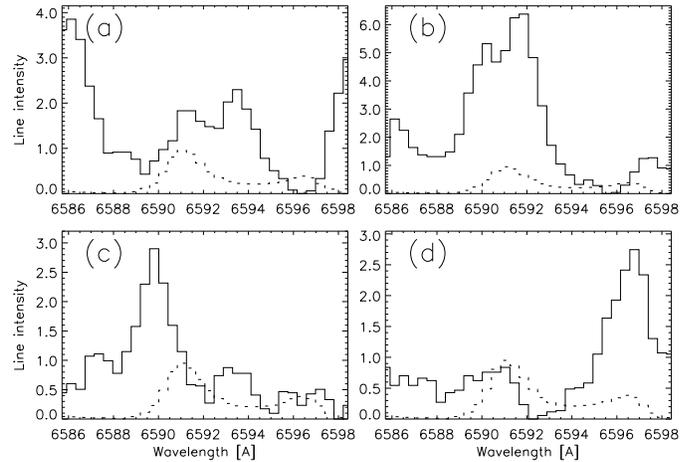}}
 \caption{
The $H_\alpha$ and [NII] profiles (solid lines) in comparison with the night sky
spectrum (dashed lines) on the same  intensity scale. {\bf a}
-- the line profile from the Region A (red ``abnormal'' component of
$H_\alpha$), {\bf b} -- the line profile from the Region F (blue ``abnormal''
component  ),  {\bf c}  --  the line profile from the Region H (only the [NII]
line without $H_\alpha$), and {\bf d} --  the $H_\alpha$ profile from the opposite
side of the galaxy.
 }
\label{figoh}
\end{figure}

Let us note that all anomalies in the emission lines profiles cannot
result from the errors of the night  sky  subtraction.
In  Fig. \ref{figoh} we present examples of the abnormal emission
profiles from Fig. \ref{figprof} and the night sky spectrum from Fig.
\ref{figfilter}b on the  same intensity scale. Fig.
\ref{figoh} a-c show some profiles with double-horned $H_\alpha$
line and/or abnormal [NII]$/H_\alpha$ ratio. In contrast, in Fig.
\ref{figoh}d we plot a ``normal`' $H_\alpha$ profile from the SW side of the
galaxy, opposite to the ``spur'' region. Obviously  all lines from
the object are more intense than the sky spectrum. Moreover the
brightest lines of the sky spectrum are  located only near the 'normal'
component of the H$_\alpha$ line (Fig. \ref{figoh}a and \ref{figoh}b). Therefore
the  ``abnormal''  component  of the $H_\alpha$ line and the largest [NII]
lines  are  not  related with overestimation or underestimation
of the sky spectrum contribution.

To summarise, the residual velocity distribution looks rather
complex. From Table \ref{spur} it can be found that the ``normally'' rotating
gas does not show a systematic deviation (within $7-10\km$) from the
line-of-sight component of circular rotation. The
``abnormal'' component of $H_\alpha$ is strongly redshifted everywhere
except  the isolated region F where the  residuals have the same order of
magnitude, but are negative.
Velocity profiles of [NII] unlike $H_\alpha$ reveal only one
component, excluding the   region E where there is a
hint that some profiles are double-horned. The velocities measured
from the [NII] profiles exceed those obtained from the ``normal''
$H_\alpha$ profile components by  $10-50\km$ in all regions with
the exception of  the Region H where the sign of the difference is
opposite. Finally, the residual velocities found from the
``abnormal'' $H_\alpha$ components and from the [NII] lines have
the same sign in all regions except the region F, which support
the hypothesis  that these velocity anomalies  could be related
phenomena.

\section{Discussion}

As  shown above, there exists at least
two different systems of emitting gas in the ``spur'' of NGC~1084.
One of the gaseous systems shows nearly circular rotation, while the
velocities of the other one differ by about $\pm100\km$ from those expected
in the case of non-perturbed rotation, and in the most  regions the
difference is positive.

The common explanation of double-peaked emission lines in star-forming
galaxies, namely an expanding superbubble, is inapplicable in this
particular case because the regions with velocity anomalies have an
unusual location in the sky plane.
As  shown in the section 2.2 the SE half of the galaxy
disk is the nearest to the observer.
 In this case the region of redshifted residual velocities  lies  on the nearest side
while the  region of blueshifted   velocities lies on the far side of the disk.
This would rather suggest a shrinking superbuble, which  is
improbable.

If a sound velocity in the gas clouds of the  ``spur'' is close to
its  mean value for the ISM in outer disk regions of spiral
galaxies (solar neighbourhood as example), $10-15\km$, then ,
the velocity of perturbed motions exceeds strongly
the sound velocity ( of  neutral hydrogen). The observed enhancement of
the [NII] emission line and the lower velocity difference measured from
the [NII] profiles may be naturally  explained if one considers  the
strong [NII] emission line as emitted by shock-excited gas slowed down
by collisions with the unperturbed medium.

Such fast-moving gas cannot be retained by the
galactic plane and must fill a volume with some filling factor,
so there could be different gas velocities along the line-of-sight.
This accounts for the complex shape of emission-line profiles
on a scale of 200pc (a pixel of 2\arcsec~ ). So it is not
surprising that the residual velocity field appears  complex being
projected onto the galactic plane. Bright HII regions,
connected with  massive gas clouds have quite normal velocities.
The perturbed component, if it exists there, might be hidden by
the bright background.

There are at least two possible interpretations of  the observed
velocity peculiarities.

One is an infall of high latitude gas clouds
(intergalactic clouds or clouds expelled from the disk earlier) onto
the galactic disk. Massive starforming clouds in the disk are
too heavy to be pushed by gas inflow, so their velocities remain
circular as  observed, although the interaction with infalling
gas may trigger star formation there. A similar phenomenon on
smaller scales is occurring in our Galaxy, where
high-velocity clouds (HVCs) are observed far from the galactic
plane. Their infall onto the  disk is confirmed by the detection of
X-ray emission from heated gas spots (Kerp et al. \cite{kerp94};
Kerp et al. \cite{kerp99} ). The existence of gas off the galactic
plane has  also been noticed in external spiral galaxies, for
example in NGC~891 ( Swaters et al. \cite{swat}).

However, invoking  the HVCs  explanation  is in contradiction with
the measure of  the  current  star formation rate (SFR) in  NGC
1084, as   deduced  from the integrated intensity of $H_\alpha$: by using the
total $H_\alpha $ flux emission  from  Kennicutt \& Kent
(\cite{kenken}) and $L_{H_\alpha}=L_{H_\alpha+[NII]}/1.5$ one gets
$L_{H_\alpha}=4.5\cdot10^{41}\,\mbox{erg~s}^{-1}$ (absorption
$A_{H\alpha}$ was accepted to be $1^{\mbox{m}}$).  The model
dependence between $L_{H_\alpha}$ and the SFR (Kennicutt \cite{ken})
gives  $\mbox{SFR}\approx 4 M_\odot\,\mbox{year}^{-1}$ for the stellar
mass  interval $0.1-100\,M_\odot$. This value is  normal (or mildly
enhanced)  for  a late-type galaxy. The ``spur'' luminosity
fraction is about of 18\% of the total $H_\alpha$ flux, and
corresponds to as  $\mbox{SFR}\approx 0.8 M_\odot\,\mbox{year}^{-1}$
in the ``spur'' region.
On the other hand, the present-day SFR  may be more intense in this galaxy:
three supernovae have been detected  during the last 40
years. These are SN 1963P , SN 1996an and SN 1998dl. The latter
two  are Type II supernovae (Nakano \cite{nakano}; Filippenko
\cite{fil}) connected with recent star formation.
The location of both  SNe II are marked by   white crosses in
Fig. \ref{figim}a. So it cannot be excluded that the observed
off-disk gas may be expelled during  the short and intense burst of
starformation which took place in this region of the galaxy.

Moreover the central parts of the HII regions in the ``spur'' has a very blue color
( the color index $(V-I_C)\approx0.4-0.5$). This  may indicate a
large fraction of  young OB stars (see Fig. \ref{figvi}).
The  red features on the E-side of the ``spur'' (Region A) with
 $(V-I_C)\approx1.-1.2$) can  be explained by  strong dust absorption
 due to  the shock waves in the ``spur'' region.

The second possible interpretation is an interaction with a gas-rich dwarf
galaxy accompanied by tidal disruptive merging. Indeed, on  the
opposite side of the galaxy, at $R\approx70''$ to the S of the
nucleus, there is a small ``island'' of $H_\alpha$ emission (Fig.
\ref{figim}a).
It does not distinguish itself dynamically, but the radio map
in the non-thermal continuum at 1.49 GHz obtained at the VLA (Condon
\cite{condon} )
shows that a long radio tail begins here which connects NGC~1084
to another radio source located 3\farcm5 (about two optical diameters)
from the galaxy. As no HI map at 21~cm is available
for NGC~1084, one cannot confirm whether  this radio tail contains some
expelled gas. But the  configuration resembles a tidal tail
as usually  developing on the opposite side of a galaxy
colliding with another one. Therefore, the  gas flow twisted in
the northern half of NGC~1084 might be accretion; if the
initial rotation momentum of this gas is nearly orthogonal to the
rotation momentum of NGC~1084, the gaseous flow could look like an
off-center polar ring. This hypothesis would explain easily the
different signs of velocity anomalies in the NW and
SE ends of the ``spur'': we would be seeing the
receding and approaching parts of the rotating "polar ring". Such
a configuration is short-lived because all the gas must fall
towards the center of the galaxy in some $10^8$ years, but this
time is not too short to prevent its detection.

To clarify further  the possible mechanisms responsible for the peculiar
velocity field of the ``spur'', high-resolution observations
of neutral and molecular gas distributions and
two-dimensional spectral investigations in various forbidden
optical emission lines are required.

\section{Conclusions}

The 2D velocity field of NGC~1084 obtained with the scanning Fabry-Perot
interferometer   tuned at the redshifted $H_\alpha$ and [NII]$\lambda 6583$
emission lines has revealed an extended region in the NE part of
the galaxy where two kinematically distinct gas systems are present.
The first system is related to normally rotating gas, whereas the second
one reveals line-of-sight velocities differing by $\pm(100-150)\km$.
The ``spur'' covers  a region  of $1-2$ kpc in size and  avoids the bright
HII regions. The fast-moving gas is mostly characterized by
enhanced [NII]$\lambda$6583 emission, which  evidences for the
presence of strong shock waves. Possible explanations for this phenomenon
are the infall of extraplanar high-velocity  clouds or the tidal
disruption and  accretion
of another small galaxy.

\begin{acknowledgements}  I would like to thank astronomers of the
Special   Astrophysical   Observatory   --   V.L.   Afanasiev and
S.N.Dodonov   for  the  observations  with FPI and for
useful discussion.  I thank as well  V.H.Chavushyan
and  V.V.Vlasyuk  who obtained   optical images at the 1m telescope and
S.V.  Drabek  who  supported observations at the 6m telescope with
the  IPCS.  I am very grateful to O.K. Sil'chenko and A.V.Zasov for help
and discussion, J. Boulesteix who provided
interference  filters  to be  used  at  the  6m  telescope  and A.N.
Burenkov   for   providing   unpublished   long-slit observations
of  NGC  1084. Also I am grateful to referees for more notes and
corrections.

This work was supported by the grant
of   the  Russian  Foundation  for  Basic  Research  (project  No.
98-02-17102).

\end{acknowledgements}

\end{document}